\documentclass[10pt,twoside]{hsqcd}
\usepackage{epsf,amsmath}
\usepackage{graphicx}

\begin{document}

\onecolumn

\title{BEAUTY \& CHARM PHYSICS AT HERA}

\author{M.~A.~BELL\\ \\
{\it for the ZEUS and H1 Collaborations} \\ \\
Oxford University\\
Denys Wilkinson Building, Keble Road, Oxford, OX1 3RH, UK\\
E-mail: m.bell1@physics.ox.ac.uk }

\maketitle

\begin{abstract}
\noindent The most recent beauty and charm results from the ZEUS and H1 collaborations are presented.
\end{abstract}

\markboth{\large \sl M.A. Bell
\hspace*{2cm} HSQCD 2005} {\large \sl \hspace*{1cm} BEAUTY \& CHARM PHYSICS AT HERA}

\section{Introduction}

Heavy quark production in electron-proton collisions at HERA is dominated by Boson Gluon Fusion at leading order 
($\gamma \mathrm{g} \to \mathrm{q}\overline{\mathrm{q}}$) in which a virtual photon coming from the electron vertex
interacts with a gluon from the proton to produce a heavy quark pair.  There are two main kinematic regimes,
Deep Inelastic Scattering (DIS) for photon virtualities Q$^2>1\;$GeV$^2$ and photoproduction ($\gamma$p) for Q$^2\simeq 0\;$GeV$^2$.
Studying the production of heavy quarks enables comparison of the experimental results to theoretical Next to Leading Order
(NLO) Quantum Chromodynamics (QCD) calculations. The experimental results can also be compared to Leading Order (LO) 
plus parton shower Monte Carlo (MC) programmes.  

\section{Charm}

Measuring the D$^*$ cross section at low Q$^2$ is a test of the NLO calculation for charm production in the 
transition region from DIS to $\gamma$p.  Q$^2$ values in the range $0.05<\mathrm{Q}^2<0.7\;$GeV$^2$ 
are reached by measuring the scattered electron in the Beam Pipe Calorimeter (BPC) \cite{zbpc}.
The measured cross section is well described by the predictions of NLO QCD (figure 1) showing that this kinematic
region is well understood theoretically.

\begin{figure}[!htb]
\vspace*{5cm}
\begin{center}
\includegraphics{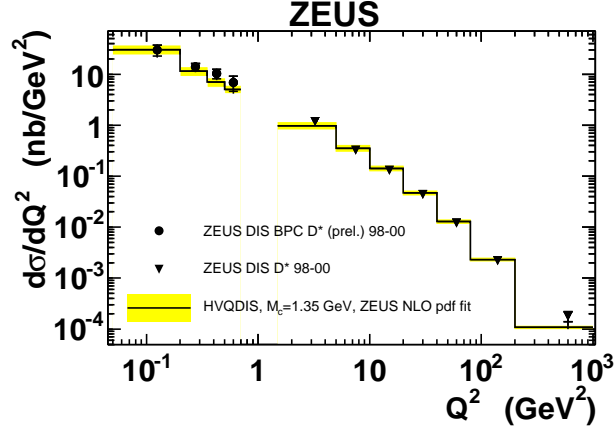}
\caption[*]{Differential D$^*$ cross section as a function of Q$^2$.}
\end{center}
\label{figa}
\end{figure}

\begin{figure}[!htb]
\vspace*{3.8cm}
\begin{center}
\includegraphics{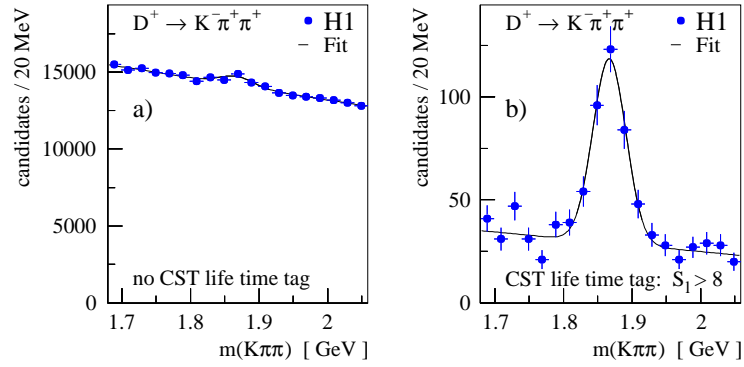}
\caption[*]{Invariant mass distributions m(K$\pi\pi$) for D$^+ \rightarrow K^-\pi^+\pi^+$ decay candidates (left) before and (right)
after a cut on the decay length significance S$_l>8$.}
\end{center}
\label{figb}
\end{figure}

Charm mesons have long lifetimes enabling them to be tagged via their displaced secondary vertices \cite{h1dimp}.
Selecting tracks depending on their decay length significance (S$_\mathrm{l}$) greatly improves the purity of the signal.
For S$_\mathrm{l}>8$, the D$^+$ signal to background ratio improves by factor of 50, and 20\% of the signal is kept 
(figure 2).

For events containing a D$^*$ and two jets in $\gamma$p, correlations can be studied between the jets
to allow detailed comparisons with QCD calculations \cite{zdstjet}.
x$_{\gamma}^{\mathrm{obs}}$ is the fraction of the photon's four momentum, manifest in the two highest p$_{\mathrm{T}}$ jets,
entering the hard dijet subprocess.
The NLO calculation describes shape of the data for direct $\gamma$p (x$_{\gamma}^{\mathrm{obs}}>0.75$), 
with the data favouring a lower charm mass.  The shape is not as well described by the NLO calculation for resolved 
$\gamma$p (x$_{\gamma}^{\mathrm{obs}}<0.75$), indicating a need for higher order corrections to the NLO calculations.
The LO + parton shower MCs particularly HERWIG fit the data well for both direct and resolved $\gamma$p (figure 3).
Similar results have been found looking at correlations between the D$^*$ and a jet not containing the D$^*$ \cite{h1dstjet}.

\begin{figure}[!htb]
\vspace*{9cm}
\begin{center}
\includegraphics{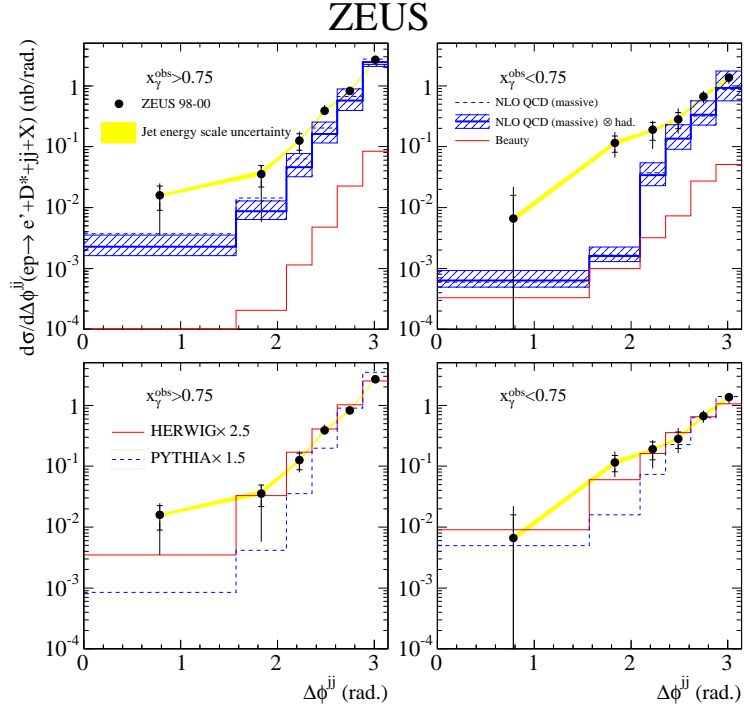}
\caption[*]{Cross section d$\sigma/$d$\Delta\phi^{jj}$ for dijets in D$^*$ photoproduction.} 
\end{center}
\label{figc}
\end{figure}

\vspace{-1cm}

\section{F$_2^{\mathrm{c}\overline{\mathrm{c}}}$ \& F$_2^{\mathrm{b}\overline{\mathrm{b}}}$ from Impact Parameters}

Using the impact parameter significance of tracks, charm and beauty fractions can be calculated
by fitting distributions in different x-Q$^2$ intervals \cite{H1f2l,H1f2h}.  Differential cross sections can then be measured
and the structure functions F$_2^{\mathrm{c}\overline{\mathrm{c}}}$ and F$_2^{\mathrm{b}\overline{\mathrm{b}}}$ evaluated from
the expression
\begin{eqnarray*}
\frac{\mathrm{d}^2\sigma^{c\bar{c}}}{\mathrm{d}x\mathrm{d}Q^2}=
\frac{2\pi\alpha^2}{xQ^4}\left[\left(1+\left(1-y\right)^2\right)F_2^{c\bar{c}}-y^2F_L^{c\bar{c}}\right]
\end{eqnarray*}
(and similarly for F$_2^{\mathrm{b}\overline{\mathrm{b}}}$) and plotted in figure 4.  
The QCD calculations fit the data reasonably well with scaling violations apparent at low x.

\begin{figure}[!htb]
\vspace*{7cm}
\begin{center}
\includegraphics{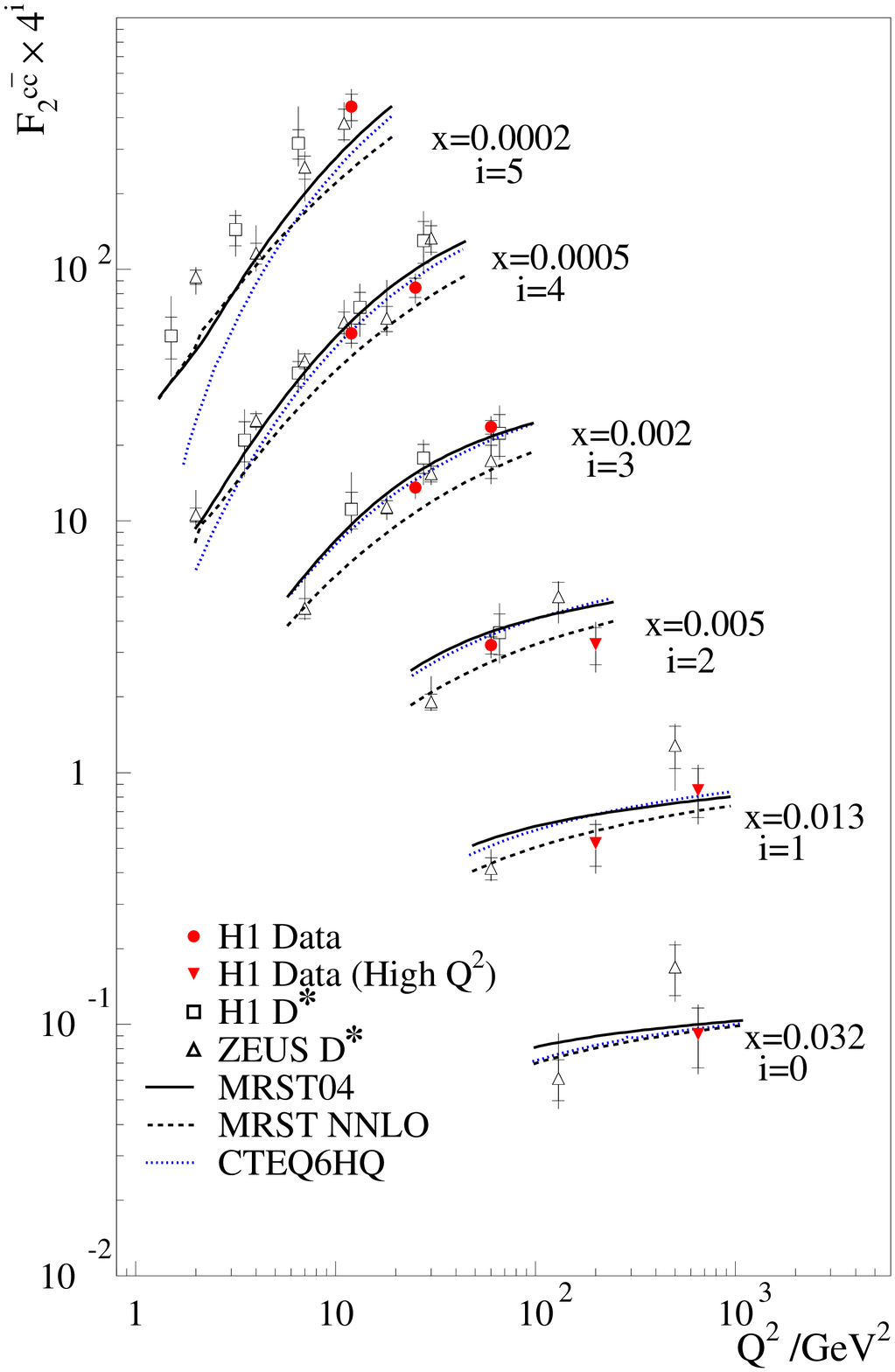}
\includegraphics{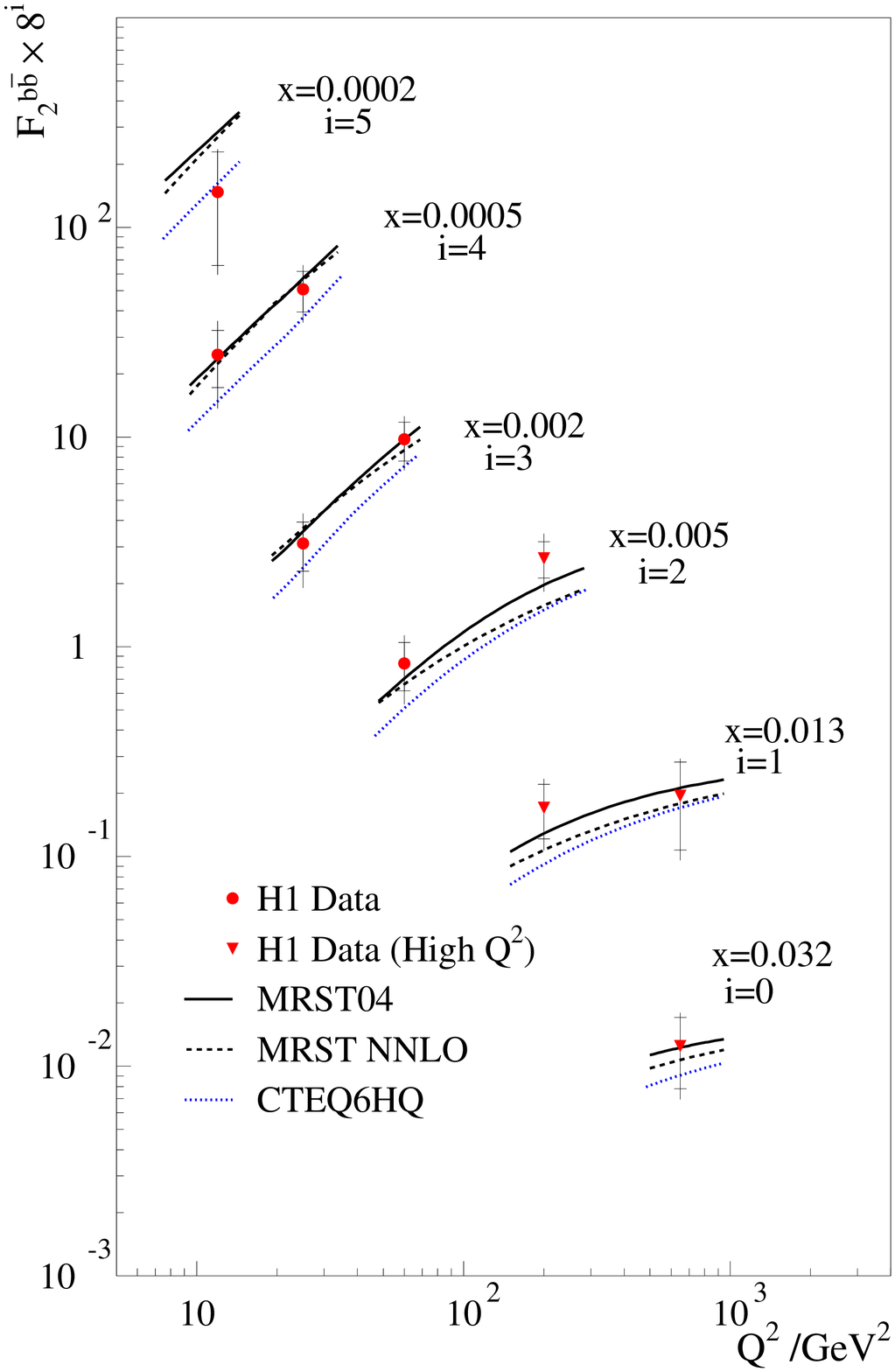}
\caption[*]{The measured $F_2^{c\bar{c}}$ and $F_2^{b\bar{b}}$ shown as a function of Q$^2$ for various $x$ values.}
\end{center}
\label{figd}
\end{figure}

\vspace{-1cm}

\section{Beauty}

Using the semileptonic decay of a B meson to a muon with an accompanying jet, the beauty can be separated from charm 
and light flavours by exploiting the high mass and long lifetime of B mesons.
By simultaneously fitting the impact parameter and the relative transverse momentum of the muon to the axis of
the associated jet the beauty fraction can be extracted \cite{zmujet,h1mujet}.
The NLO calculation describes the data reasonably well, though H1 have excess of events at low p$_\mathrm{T}^{\mu}$
(figure 5).

By selecting events with two muons the background from charm and light flavours is suppressed.
Separating the sample into high and low mass, isolated and non-isolated, like and unlike sign regions
further constrains the background \cite{zmumu} and although this results in low statistics, the data agree with the
theory and MC predictions (figure 6).

Figure 7 displays a summary of the beauty results from HERA showing a good coverage of measurements.  There
is a tendency of the data to lie above the NLO prediction though measurements with smaller errors are closer to the theory.
Improved theoretical understanding is needed to include higher orders but also improved precision from the experimental
results.

\begin{figure}[!htb]
\vspace*{5.0cm}
\begin{center}
\includegraphics{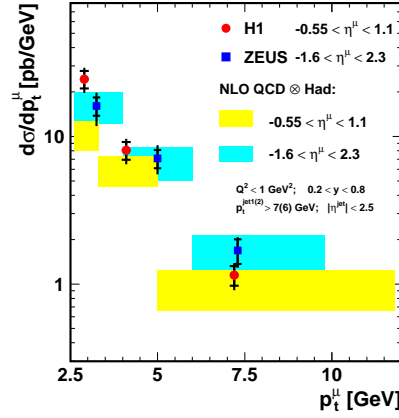}
\caption[*]{Differential cross section d$\sigma/\mathrm{d}p_T$ for the $\gamma$p process 
$ep\rightarrow eb\bar{b}X\rightarrow ejj\mu X^{\prime}$.}
\end{center}
\label{fige}
\end{figure}

\begin{figure}[!htb]
\vspace*{3.1cm}
\begin{center}
\includegraphics{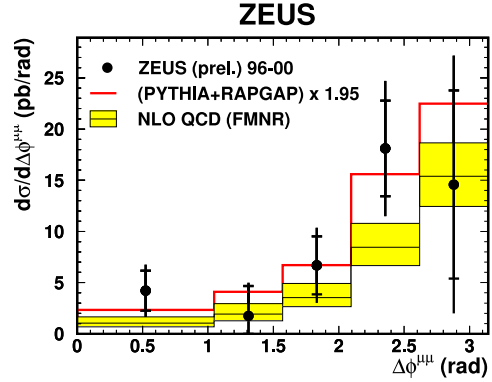}
\caption[*]{Cross section $d\sigma/d\Delta\phi^{\mu\mu}$ for dimuon events from $b\bar{b}$ decays in which each muon originates
from a different $b(\bar{b})$ quark.}
\end{center}
\label{figf}
\end{figure}

\begin{figure}[!htb]
\vspace*{6.0cm}
\begin{center}
\includegraphics{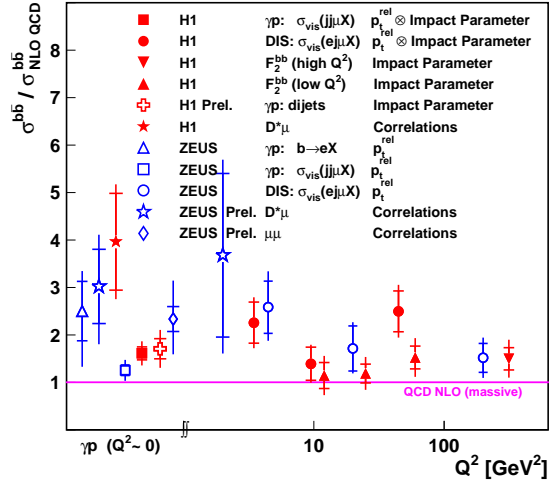}
\caption[*]{Ratio of beauty cross sections to the NLO QCD prediction for different measurements from HERA.}
\end{center}
\label{figg}
\end{figure}

\vspace{-1.5cm}

\section{Conclusions}

In general the charm results from HERA are in good agreement with NLO QCD.  The beauty results are not as precise
but tend to indicate that higher order theory calculations are needed.

\end{document}